\documentclass[letter]{aa} 
\usepackage{graphicx}
\usepackage{txfonts}
\usepackage{natbib,twoopt}
 \usepackage[breaklinks=true]{hyperref} 
 \bibpunct{(}{)}{;}{a}{}{,}    
 \newcommandtwoopt{\citeads}[3][][]{\href{http://adsabs.harvard.edu/abs/#3}%
                                        {\citealp[#1][#2]{#3}}}
 \newcommandtwoopt{\citepads}[3][][]{\href{http://adsabs.harvard.edu/abs/#3}%
                                        {\citep[#1][#2]{#3}}}
 \newcommandtwoopt{\citetads}[3][][]{\href{http://adsabs.harvard.edu/abs/#3}%
                                        {\citet[#1][#2]{#3}}}
 \newcommandtwoopt{\citeyearads}[3][][]%
   {\href{http://adsabs.harvard.edu/abs/#3}{\citeyear[#1][#2]{#3}}}

\begin{document}

   \title{Simultaneous optical and near-infrared linear spectropolarimetry of the earthshine}
 \author{P. A. Miles-P\'aez
          \inst{1}\fnmsep\inst{2}
          \and
           E. Pall\'e\inst{1}\fnmsep\inst{2}
          \and
          M. R. Zapatero Osorio\inst{3}
          }

   \institute{Instituto de Astrof\'\i sica de Canarias, La Laguna, E38205 Spain\\
              \email{pamp@iac.es; epalle@iac.es}
         \and
             Departamento de Astrof\'\i sica, Universidad de La Laguna, Av$.$ Astrof\'\i sico Francisco S\'anchez, s$/$n E38206--La Laguna, Spain\
          \and
             Centro de Astrobiolog\'\i a (CSIC-INTA), Carretera de Ajalvir km 4, E-28850 Torrej\'on de Ardoz, Madrid, Spain\\
             \email{mosorio@cab.inta-csic.es}
             }

  \date{Received 2013; accepted 2013}

 
  \abstract
   {}
   {We aim to extend our current observational understanding of the integrated planet Earth spectropolarimetry from the optical to the near-infrared wavelengths. Major biomarkers like O$_{\rm 2}$ and water vapor are strong flux absorbents in the Earth atmosphere and some linear polarization of the reflected stellar light is expected to occur at these wavelengths.}
   {Simultaneous optical ($0.4-0.9$ $\mu$m) and near-infrared ($0.9-2.3$ $\mu$m) linear spectropolarimetric data of the earthshine were acquired by observing the nightside of the waxing Moon. The data have sufficient spectral resolution (2.51 nm in the optical, and 1.83 and 2.91 nm in the near-infrared) to resolve major molecular species present in the Earth atmosphere. }
   {We find the highest values of linear polarization ($\ge 10\%$) at the bluest wavelengths, which agrees with the literature. Linear polarization intensity steadily decreases towards red wavelengths reaching a nearly flat value beyond $\sim$0.8 $\mu$m. In the near-infrared, we measured a polarization degree of $\sim$4.5\,\%~for the continuum. We report the detection of molecular features due to O$_{2}$ at $0.760, 1.25\,\mu$m and H$_{2}$O at 0.653--0.725\,$\mu$m, 0.780--0.825\,$\mu$m, 0.93 and 1.12\,$\mu$m in the spectropolarimetric data; most of them show high linear polarimetry degrees  above the continuum. In particular, the broad H$_{2}$O 1.12\,$\mu$m band displays a polarimetric intensity as high as that of the blue optical. These features may become a powerful tool to characterize Earth-like planets in polarized light. }
   {}


   \maketitle
\section{Introduction}
The increasing progress in instrumentation and observational techniques has allowed the detection of over 1,000 planets with masses ranging from a few times the Jupiter mass to the super-Earth regime ($\lesssim10$\,M$_{\bigoplus}$). The most recent observational studies (e.g., \citeads{2012Natur.481..167C}) suggest that as many as 62$\%$ of the Milky Way stars can harbor Earth-like planets. It seems probable that a significant number of Earth-sized planets located within their parent stars habitable zone will be found in the near future. 

Transit spectroscopy is currently the most extended technique to characterize the planetary atmospheres of the new discoveries, and has proved successful for gaseous giant planets (e.g., \citeads{2008Natur.452..329S}). The atmospheres of Earth-sized planets are too difficult to study with actual instrumentation. While biomarkers are readily identifiable in the Earth's transmission spectrum \citepads{2009Natur.459..814P}, even in the most favorable case of an Earth-like planet orbiting small stars or brown dwarfs, several transits would be necessary for a reliable detection even using large-aperture (and space-based) telescopes \citepads{2009ApJ...698..519K,2011ApJ...728...19P}. The probability of finding a transiting exo-Earth in the habitable zone of a G2V-type star is $\sim0.5\,\%$~\citepads{2012ApJS..201...15H}. The direct detection of the light reflected/emitted by the exo-Earth would be necessary for its proper characterization; this is indeed a very challenging task. 

Polarization may contribute to the analysis of the atmospheres of these new worlds by taking advantage of the fact that the stellar light is generally not polarized, whereas the light reflected from the planet is. Various groups \citepads{2008A&A...482..989S,2011A&A...530A..69K} have theoretically studied the optical linear polarimetric signal of an Earth-like planet as a function of orbital phase and wavelength. Their predictions must be tested against observations of the only `Earth-like planet' known so far: the Earth. This can be done by observing the earthshine, which is the sunlight scattered by the dayside Earth and reflected by the nightside of the Moon. The optical and near-infrared earthshine spectra reveal atmospheric and surface biosignatures and are sensitive to features like water clouds, oceans, deserts, and volcanic activity (e.g., \citeads{2002ApJ...574..430W,2006ApJ...644..551T,2006ApJ...651..544M,2009Natur.459..814P,2012ApJ...755..103G}); many are present in the earthshine optical spectropolarimetric data \citepads{2012Natur.483...64S, 2013PASJ...65...38T}.

Here, we provide the first linear spectropolarimetric measurements of the earthshine from the  visible to the near-infrared (NIR) wavelengths (0.4--2.3 $\mu$m). We used observations of the Moon's dark side  for a characterization of the integrated polarimetric properties of the Earth. These observations  extend our current understanding of the optical linear polarimetric data of the earthshine  (\citeads{2013A&A...556A.117B} and references therein) towards red wavelengths. 


\section{Observations and data analysis}\label{data}

Earthshine optical and NIR  linear polarimetric spectra were taken using the  Andaluc\'\i a faint object spectrograph and camera (ALFOSC) and the long-slit intermediate resolution infrared spectrograph (LIRIS; \citeads{2004SPIE.5492.1094M}) of the 2.56-m Nordic Optical Telescope (NOT) and the 4.2-m William Herschel Telescope (WHT), respectively. Observing date was 2013 May 18; the journal of the observations is provided in Table~\ref{table1}.

ALFOSC polarimetric mode uses a calcite plate that provides simultaneous measurements of the ordinary and the extraordinary components of two orthogonally polarized beams separated by 15\arcsec, and a half-wave plate that we rotated four times in steps of 22.5$^{\circ}$. For spectroscopy we employed the grism $\#$4 and a slit 1\farcs5 width and $\sim$15\arcsec long, which in combination with a pixel pitch of 0\farcs19 (2048\,$\times$\,2048 E2V detector) yield a wavelength coverage of $\sim 0.4-0.9\,\mu$m and a spectral resolution of 2.51 nm ($R \sim 250$ at 0.65 $\mu$m). 

LIRIS polarimetric mode uses a wedged double Wollaston device \citepads{1997A&AS..123..589O}, consisting of the combination of two Wollaston prisms that deliver four simultaneous images of the polarized flux at vector angles 0$^{\circ}$ and 90$^{\circ}$, 45$^{\circ}$ and 135$^{\circ}$. An aperture mask 4\arcmin$\,\times\,$1\arcmin~in size is in the light path to prevent overlapping effects between the different polarization vector images. Two retarder plates provide polarimetric images with exchanged orthogonal vector angles, useful to minimize flat-fielding effects and provide better signal-to-noise (S/N) data \citepads{2011ApJ...740....4Z,2013A&A...556A.125M}. For spectroscopy we employed the grisms $ZJ$ and $HK$, and a slit 0\farcs75 width and 50\arcsec~long, which in combination with a pixel size of 0\farcs25 (1024\,$\times$\,1024 HgCdTe HAWAII array) yield  wavelength coverages of $\sim0.9-1.51\,\mu$m ($ZJ$) and $\sim1.4-2.4\,\mu$m ($HK$), and spectral resolutions of 1.83 nm ($R \sim 540$ at 1 $\mu$m) and 2.91 nm ($R \sim 540, 750$ at 1.6 and 2.2 $\mu$m).

\begin{figure}
   \begin{center}
    \includegraphics[width=0.31\textwidth]{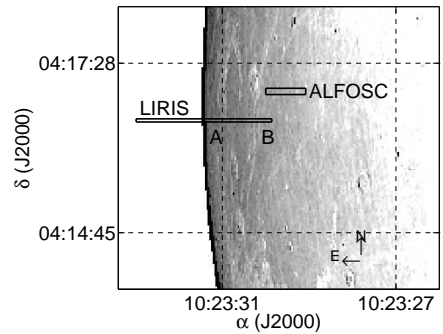}
     \caption{Relative positions of the ALFOSC and LIRIS slits on the Moon surface during the observations of the lunar nightside on 2013 May 18. Equatorial coordinates $\alpha$ and $\delta$ correspond to 21$^{\rm h}$ 20$^{\rm m}$ universal time. For displaying purposes, the slit widths are multiplied by a factor of four, and the slit lengths are plotted on true scale.}
              \label{Fig1}
     \end{center}
    \end{figure}

Spectropolarimetric observations were carried out under photometric sky conditions and raw seeing of 0\farcs9. The angle formed by the Sun, the Moon, and the Earth was 81\degr, and the waxing Moon illuminated area was 59\%. At the beginning of the astronomical night, the NOT and WHT were pointed to the center of the lunar disk, and an offset was applied to move both telescopes to the dark side (East) of the Moon. To track the Moon with the NOT during the observations, we manually introduced proper right ascension and declination rates\footnote{Ephemeris computations by the Jet Propulsion Laboratory HORIZONS project: http://ssd.jpl.nasa.gov/?horizons} every five minutes. The location of the ALFOSC and LIRIS slits on the Moon is illustrated in Figure~\ref{Fig1}. Half of the LIRIS slit was positioned at the Moon terminator and the ``dark'' side while the other half registered the sky contribution simultaneously. Because of the small size of the ALFOSC slit, alternate Moon and sky observations were conducted at a separation of 600\arcsec. 

One polarimetric cycle consisted of four consecutive frames with half-wave plate positions of 0\degr, 22\fdg5, 45\degr, and 67\fdg5 (ALFOSC), and two consecutive frames obtained with two different retarder plates (LIRIS). Typical exposure times per frame were 90 s and 400 s for the optical and NIR; one full polarimetric cycle was thus completed in about 7 min (ALFOSC) and 14 min (LIRIS) including 0.5--1-min overheads. We collected a total of 15 cycles of earthshine polarimetric data and 7 cycles of ``sky'' observations with ALFOSC, and 6 ($ZJ$) and 4 ($HK$) cycles with LIRIS. Before sky subtraction, all LIRIS frames were rectified using the telluric emission lines and observations of Ar and Xe arcs so that the spatial axis lies perpendicular to the spectral axis. The earthshine ALFOSC data were sky subtracted using sky frames obtained very close in time and typically within $\pm$0.1 airmass. Wavelength calibration was performed using observations of He$+$Ne (ALFOSC) and Ar$+$Xe (LIRIS) lamps, with typical uncertainty of $\pm$0.15--0.20\,\AA. 
%

The normalized Stokes parameters $q$ and $u$ were computed using the flux ratio method and the equations given in \citetads{2005ApJ...621..445Z} for the optical and in \citetads{2011ApJ...740....4Z} for the NIR. We obtained as many $q$ and $u$ spectra as polarimetric cycles were observed at optical wavelengths. To improve the signal-to-noise (S/N) ratio of the data, we present here final spectra computed by combining all polarimetric cycles and by collapsing all pixels along the spatial direction. A total of 65 ALFOSC pixels (or 12\farcs35) per wavelength were fused together into one-dimension spectra, while LIRIS data were folded up into two regions (indicated in Figure~\ref{Fig1}): one (A, 39 pix or 9\farcs75) close to the Moon terminator, and another (B, 38 pix or 9\farcs5) spatially near the location of the ALFOSC slit; each section corresponds to $\sim$1/4 of the total LIRIS slit length. In the collapsing processes, we obtained the average of all pixels and rejected the maximum and minimum values per wavelength. Optical data have better S/N than NIR spectra by factors of $\sim$15 (optical versus $ZJ$) and $\sim$25 (optical versus $HK$). The linear polarization degree was derived as the quadratic sum of the normalized Stokes parameters. The debiased polarization degree, $p^{\ast}$, was computed by considering the polarization uncertainty (see below) and according to the equation given in \citetads{1974ApJ...194..249W}. 

Together with the earthshine data, observations of polarized and non-polarized standard stars (HD\,154892, HD\,161056, and Elia\,2-25) from the catalogs of \citetads{1992AJ....104.1563S} and \citetads{1992ApJ...386..562W} were acquired with the same instrumental configuration and on the same observing night as the earthsine, and were used to control the instrumental linear polarization and to check the efficiency of the ALFOSC and LIRIS optics. The same unpolarized standard was targeted at the two sites. We found an upper limit of 0.1\%~on the instrumental linear polarization degree across the explored wavelengths. We also found that the measured vibration angle of the linear polarization, $\Theta$, has to be corrected by a constant value ($+4\fdg3 \pm 1\fdg3$) at the LIRIS $Z$, $J$, $H$, and $K$ wavelengths (which agrees with the assumption made by \citeads{2013A&A...556A.125M}), while the angle correction is wavelength-dependent for ALFOSC: from $-2\fdg0$ at 0.4 $\mu$m to $+5\fdg0$ at 0.9 $\mu$m. We used the observations of the zero-polarized standard stars (conveniently scaled in terms of S/N) to estimate the uncertainties associated with the earthshine $q$, $u$, and $p^{\ast}$ measurements. They are as follows for $p^{\ast}$: $\pm$0.3\%~at 0.45 $\mu$m, $\pm$0.2\%~at 0.60 $\mu$m, $\pm$0.6\%~at 0.85 $\mu$m, $\pm$0.7\%~at 1.04 $\mu$m, $\pm$0.6\%~at 1.25 $\mu$m, $\pm$0.8\%~at 1.64 $\mu$m, and $\pm$1.2\%~at 2.15 $\mu$m.

\begin{table*}
\caption{Observing log (2013 May 18).}
\label{table1}
\centering
\begin{tabular}{l c c c c c c c}
\hline\hline
Instrument&Elapsed time&Exp. time$^{\rm a}$&Airmass&$\Delta \lambda$&$\lambda$&$p^{*}$&$\Theta$\\
 &(UT)&(s) & &($\mu$m) &($\mu$m) &(\%)&(deg)\\
\hline
ALFOSC &21:20--00:32 &15$\times$4$\times$90 &1.19--2.58 &0.4--0.9&0.428 &$10.5\pm0.3$ &$112.2\pm0.8$ \\ 
  & & & & &0.535&$7.7\pm0.2$&$112.3\pm0.7$ \\ 
  & & & & &0.631&$6.4\pm0.2$&$113.0\pm1.1$ \\
  &  & & & &0.810&$5.7\pm0.6$&$114.8\pm4.9$ \\
LIRIS &21:32--23:05 &6$\times$2$\times$400&1.22--1.57&0.9--1.5&1.035&$4.0\pm0.7$&$111.8\pm4.9$\\
  &  & & & &1.250&$4.2\pm0.6$&$119.2\pm4.2$ \\
LIRIS  &23:08--00:01&4$\times$2$\times$400&1.67--2.18&1.4--2.4&1.640&$4.6\pm0.8$&$109.8\pm5.2$ \\
  &  & & & &2.150&$4.0\pm1.2$&$106.3\pm8.6$ \\
\hline
\end{tabular}
\tablefoot{
\tablefoottext{a}{ Number of polarimetric cycles as defined in Section\,\ref{data}, $\times$ number of frames in one cycle, $\times$ integration time per frame.}\\
}
\end{table*}
%
%
%

\section{Results and discussion}\label{results}

   \begin{figure}
   \begin{center}
    \includegraphics[width=0.45\textwidth]{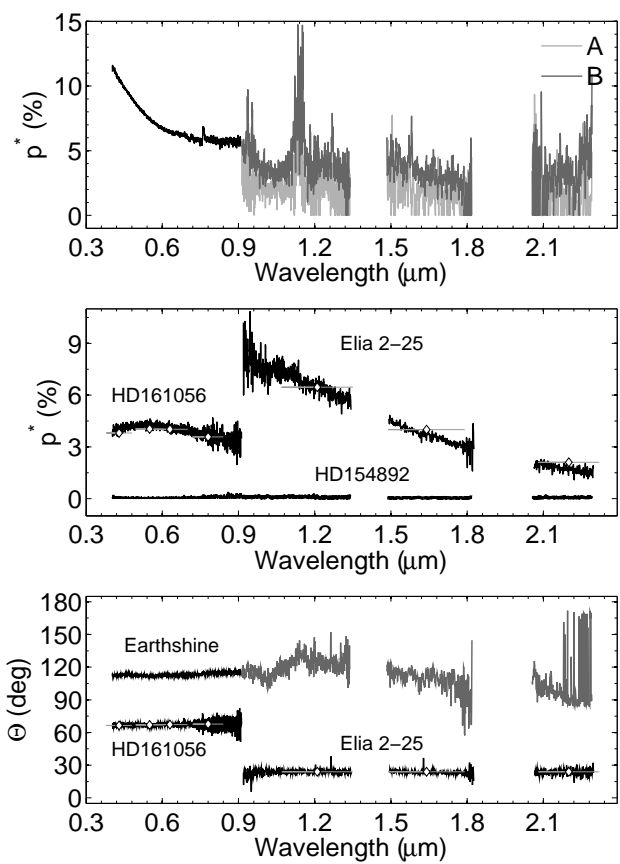}
     \caption{{\sl Top:} Spectroscopic visible (black line) and NIR (gray lines) linear polarization degree of the earthshine obtained on 2013 May 18. Region B (see text) lies spatially near the location of the optical observations; region A includes the Moon terminator. {\sl Middle:} Spectroscopic visible and NIR linear polarization degree of observed polarized and non-polarized standard stars (black lines). Literature photometric polarimetry \citepads{1992AJ....104.1563S,1992ApJ...386..562W} is plotted as white diamonds, gray horizontal error bars represent the filter widths. {\sl Bottom:} Spectroscopic polarization vibration angle of the earthshine and polarized standard stars. Symbols are as in previous panels. In all panels, wavelengths strongly affected by telluric absorption are removed. }
              \label{Fig2}
   \end{center}
    \end{figure}
%

Figure~\ref{Fig2} depicts the ALFOSC and LIRIS debiased linear polarization degree and vibration angle of polarization ($\Theta$) spectra for both the Earth and the standard stars. Note that the optical and NIR polarized stars (middle and bottom panels) are different sources. The data of the standards agree with the literature measurements at the 1-$\sigma$ level, thus supporting our confidence in the data analysis and instruments performance. 

We show the Earth NIR $p^{*}$ spectra (top panel of Figure~\ref{Fig2}) for the two Moon sections A and B of the LIRIS slit (see Figure~\ref{Fig1}). Both Moon regions yield the same spectral shape and display similar features, but the polarization intensity of region B (more similar to the optical polarimetric signal and also spatially closer to the ALFOSC slit) is a factor of $1.8\pm0.3$ larger than that of region A. This factor remains practically constant throughout all NIR wavelengths. Since regions A and B extend over $\ge$17 km each on the Moon limb, we attribute this difference to the fact of observing distinct lunar superficial regions, e.g., maria and highlands. Our finding agrees with the recent study by \citetads{2013A&A...556A.117B}; these authors measured the earthshine $p^{*}$ ($P\approx p^{*}$ since $P\ggg\sigma_{P}$) by observing simultaneously highland and mare lunar areas using optical filters, and found that mare regions yield values of $p^{*}$ a factor of 1.3 times larger than highlands. There is also an offset of about 0.5--1.0\,\%~in the linear polarimetry degree between the red wavelengths of the optical observations and the NIR B-region data, which may be also explained by the monitoring of distinct lunar areas. This highly contrasts with the excellent optical--NIR match of the polarization vibration angle measurements seen in the bottom panel of Figure~\ref{Fig2}. We provide Earth $p^{*}$ and $\Theta$ measurements (B region) at different wavelengths of reference in the last three columns of Table\ref{table1}. 

The most noticeable feature of the earthshine linear polarization degree is that the polarization intensity is the largest ($\ge 10\%$) at the bluest wavelengths and steadily decreases towards the red. Various groups \citepads{1957SAnAp...4....3D,2005ASPC..343..211W,2012Natur.483...64S,2013PASJ...65...38T,2013A&A...556A.117B} have reported a similar property. In our data, we observe that the linear polarization intensity remains nearly constant at about 5.7\,\%~between $\sim0.8\,\mu$m and $0.9\,\mu$m. Interestingly, the general shape of the NIR $1-2.3\,\mu$m polarimetric spectrum (except for some signatures due to water vapor and oxygen) also appears rather flat within the quoted uncertainties with a value around $p^{*} = 4.5$\,\%~independently of wavelength. On the contrary, the behavior of the vibration angle appears slightly different. Whereas it remains constant within $\pm3\degr$ for all optical frequencies, the broad-shaped  variations of up to 30\degr~detected in the NIR (which cannot be attributed to instrumental or measurement errors) seem to correlate with the spectral features observed in $p^{*}$. This may be attributed to the wavelength-dependent relative contribution of different components (atmospheric species, clouds, aerosols, and surface albedos) to the globally-integrated linear polarization of the earthshine. 

   \begin{figure}
    \centering
    \includegraphics[width=0.49\textwidth]{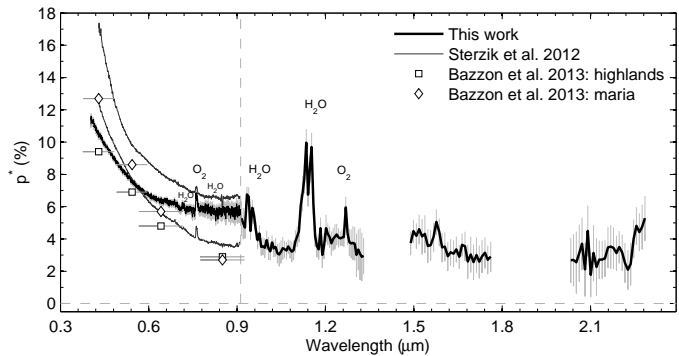}
     \caption{Our visible and NIR spectropolarimetric measurements of the earthshine are compared to literature data. A 10-pixel binning was applied to the NIR spectrum of region B. The uncertainty per wavelength is plotted as vertical gray error bars. Wavelengths of strong telluric absorption have been removed. Some molecular species seen in ``emission" (indicative of strong atmospheric flux absorption and less multiscattering processes occurring at those particular wavelengths) are labeled. The vertical dashed line separates the ALFOSC and LIRIS data. }
            \label{Fig3}
    \end{figure}

We compare our measurements with data from the literature in Figure~\ref{Fig3}. To improve the quality of the NIR linear polarization degree spectrum of region B, we applied a 10-pixel binning in the spectral dimension. Overlaid in Figure \ref{Fig3} are the optical $p^{*}$ values obtained at a spectral resolution of 3 nm and for two separated dates by \citetads{2012Natur.483...64S}, and the broadband filter measurements of Moon highlands and maria made by \citetads{2013A&A...556A.117B} for a Sun-Earth-Moon phase angle similar to ours.  All optical data display a qualitatively similar pattern (previously discussed), but they differ quantitatively in the amount of polarization per wavelength and the spectral slope. The spectral slope of \citetads{2012Natur.483...64S} and \citetads{2013A&A...556A.117B} data is steeper than the ALFOSC spectrum, while our measurements and those of \citetads[][see their Figure~3]{2013PASJ...65...38T} display related declivity. These differences may be understood in terms of distinct lunar areas explored by the various groups and different observing dates. \citetads{2012Natur.483...64S} attributed the discrepancies of their two spectra solely to the time-dependent fraction of Earth clouds, continents, and oceans contributing to the earthshine. Our simultaneous NIR spectra of regions A and B support the conclusion of \citetads{2013A&A...556A.117B} that linear $p^{*}$ values may also partially depend on the exact location of the Moon observed. 


Despite the featureless appearance of the polarimetric spectrum of the Earth, some signatures are still observable at the level of $\ge$3\,$\sigma$ and at the resolution and quality of our data. The most prominent molecular features have been identified as labeled in Figure~\ref{Fig3}: the optical O$_{2}$ at $0.760\,\mu$m (previously reported by \citeads{2012Natur.483...64S}) and H$_{2}$O in the intervals 0.653--0.725\,$\mu$m and 0.780--0.825\,$\mu$m, and the NIR H$_{2}$O at 0.93\,$\mu$m and $1.12\,\mu$m and O$_{2}$ at 1.25 $\mu$m, all the latter reported here for the first time. Linear polarization is  larger inside deep absorption molecular bands because strong opacity leaves only upper atmospheric layers to contribute significantly to the observed flux, thus reducing multiple scattered with respect to single scattered photons. As discussed by \citetads{2003ApJ...585L.155S} and \citetads{2008A&A...482..989S}, single scattering produces more intense polarization indices than multiple scattering events. The presence of the O$_{2}$ A-band at $0.760\,\mu$m and H$_{2}$O at 0.653--0.725\,$\mu$m, 0.780--0.825\,$\mu$m, and $0.93\,\mu$m in the Earth spectropolarimetry was already predicted by \citetads{2008A&A...482..989S}. These authors stressed the sensitiveness of the A-band polarization index to the planet gas mixing ratio and altitude of the clouds. Interestingly, the peak of the linear polarization at the center of the 1.12-$\mu$m H$_{2}$O band is $\sim2.7$ times larger than the values of the surrounding continuum, i.e., similar in intensity to the blue optical wavelengths. Even more important should be the linear polarization signal of water bands at $\sim1.4\,\mu$m and $\sim1.9\,\mu$m, which unfortunately cannot be characterized from the ground due to strong telluric absorption. Spectropolarimetric models of the Earth, guided by the visible and NIR observations shown here, could provide hints about their expected polarization values. 

To this point, we have presented earthshine $p^{*}$ values as measured from the light deflection on the Moon surface. However, this process introduces significant  depolarization due to the back-scattering from the lunar soil \citepads{1957SAnAp...4....3D}; the true linear polarization intensity of the planet Earth is actually larger. At optical wavelengths, the depolarization is estimated at a factor of $3.3 \lambda/550$ ($\lambda$ in nm) by \citetads{1957SAnAp...4....3D}, making true polarization be in the range 26--31\%. We are not aware of any determination of the depolarization factor for the NIR in the literature. The extrapolation of equation~9 by \citetads{2013A&A...556A.117B} towards the NIR yields corrections of $\sim \times2.2-\times3.3$ for the wavelength interval $0.9-2.3\,\mu$m, implying that the true linear polarization intensity of the Earth may be $\sim9-12\%$ for the NIR continuum, and $\sim12-36\%$ at the peak of the O$_{2}$ (1.25\,$\mu$m)  and H$_{2}$O (1.12\,$\mu$m) bands. The extrapolation of \citetads{1957SAnAp...4....3D} depolarization wavelength dependency towards the NIR would yield even higher true polarization values by a factor of $\sim$4. 
Further modeling efforts are needed to confirm these features, which may become a powerful tool for the search for Earth-like worlds and their characterization in polarized light.

\begin{acknowledgements}
Based on observations made with the William Herschel and Nordic Optical Telescopes operated on the island of La Palma by the Isaac Newton Group and the Nordic Optical Telescope Scientific Association (NOTSA), respectively, in the Spanish Observatorio del Roque de los Muchachos of the Instituto de Astrof\'\i sica de Canarias. The instrument ALFOSC is provided by the Instituto de Astrof\'\i sica de Andaluc\'\i a under a joint agreement with the University of Copenhagen and NOTSA. We thank Dr. D. Stam for helpful discussion. This work is partly financed by the Spanish Ministry of Economics and Competitiveness through projects AYA2012-39612-C03-02 and AYA2011-30147-C03-03.
\end{acknowledgements}
%
%


\bibliographystyle{aa} 
\bibliography{biblio.bib} 
\end{document}